\documentclass[12pt,epsfig]{article}   
\usepackage{epsfig}
\title{
       Selfconsistent calculations of fission barriers in the Fm region.}
\author{
  M. Warda$^{1,2}$, J. L. Egido$^1$, L. M. Robledo$^1$, and K. Pomorski$^{2,3}$
\\
     {\it $^1$Departamento de F\'\i sica Te\'orica C-XI, Universidad Aut\'onoma
     de Madrid,} \\
                           {\it Madrid, Spain}
\\
{\it $^2$Katedra Fizyki Teoretycznej, Uniwersytet M. C. Sk\l odowskiej,}\\
                           {\it  Lublin, Poland}
\\
      {\it $^3$IReS -- IN$_2$P$_3$ -- CNRS and Universit\'e Louis Pasteur,}\\
                           {\it Strasbourg, France}}

\date{}

\begin{document}

\maketitle

\noindent
\begin{abstract}

The fission barriers of the nuclei $^{254}$Fm, $^{256}$Fm, $^{258}$Fm,  $^{258}$No and
$^{260}$Rf are investigated in a fully microscopic way up to the scission
point. The analysis is based on the constrained Hartree-Fock-Bogoliubov
theory and Gogny's D1S force. The quadrupole, octupole and hexadecapole moments
as well as the number of nucleons in the neck region are used as constraints.
Two fission paths, corresponding to the bimodal fission, are found. The decrease
with isotope mass of the half-life times of heavy Fm isotopes is also explained.\\   \\
\end{abstract}
 


\section{Introduction}                                                         

Due to the loss of stability, with respect to spontaneous fission, the number
of elements is limited to only few more than a hundred and ten. Both the
experimental and theoretical studies of spontaneous-fission properties are
crucial for understanding the stability properties of the heaviest elements.
The abrupt transition that occurs from $^{254}$Fm to $^{258}$Fm in
fission-fragment mass and kinetic-energy distributions, and in the spontaneous
fission half-lives of heavy nuclei was found experimentally  (see e.g. the
review articles \cite{Hof89,Goe99}). For $^{258}$Fm and heavier isotopes, the
spontaneous-fission half-life decreases relative to $^{256}$Fm by several
orders of magnitude. The mass distribution of fission fragments of $^{258}$Fm
becomes very narrow with a single peak at symmetrical fission and the
kinetic-energy distribution has two peaks: one at high energy  (230 MeV) and
the second less prominent at lower energy (205 MeV). On the other hand,
the  $^{256}$Fm isotope exhibits a rather strong mass asymmetric distribution 
$\bar{A}_L/\bar{A}_H=112/141$ and only the low energy peak is observed in the
kinetic energy distribution. This is a rather puzzling situation as it is
not expected, from a macroscopic point of view, a substantial change in the 
properties of both isotopes. Therefore, the different fission properties of both isotopes
have to be attributed to subtle shell effects making its theoretical explanation
even more challenging.

The qualitative explanation of all these phenomena by the existence of an
additional fission valley in the multidimensional potential energy surface was
proposed by Hulet et al. \cite{Hul86,Hul89}. They assumed a bimodal character
of the kinetic-energy and mass distributions to show that for $^{258}$Fm there
should exist two different fission paths leading to two distinctly different
scission configurations. The first one is the conventional scission
configuration of two fairly elongated shapes corresponding to the low
kinetic-energy peak and the broad mass distribution. The second one leading to
compact scission, i.e. configuration of two touching spheres corresponding to
the high energy peak in the kinetic energy distribution and the narrow peak in
mass distribution at symmetric  fission.

After the discovery of bimodal fission in $^{258}$Fm \cite{Hul86,Hul89}, a
number of  theoretical papers has focused on this problem
\cite{Bro86,Moe87,Pas88,Cwi89,Moe89,Moe00,Moe01}. All these papers are based on the mean--field
single-particle potential and the Strutinsky shell correction method. Most of
them deal with the form of the potential energy surface only. These static
calculations usually give two fission  valleys: one leading to the elongated
form  of fission fragments (EF) and the second one which corresponds to two
nearly spherical fragments, which is usually referred as the compact fission
(CF) valley. It has been pointed out by Brosa \cite{Bro86} and others that this
new CF valley is associated with the doubly magic (Z=50, N=82) shell closure in
the fission fragments.

The macroscopic--microscopic calculations of the potential energy surface (PES)
for $^{258}$Fm reported in Ref. \cite{Cwi89} are based on the  Woods-Saxon
single particle Hamiltonian.   The collective potential energy surface
$V(\beta,\beta_4)$ was minimised there with respect to the deformation
parameters $\beta_3$, $\beta_5$ and $\beta_6$. Two fission valleys were found,
bifurcating right after the exit from the fission barrier. It was also shown in
\cite{Cwi89} that for the largest  values of $\beta$ ($\beta=1.6$) the EF
valley correspond again to symmetric fission $(\beta_3=\beta_5=0)$. This
theoretical result is in line with the observation of Hulet et al.
\cite{Hul86,Hul89} where they found symmetric fission only and equal fission
half-life times for both fission modes, i.e. for the fragments with the large
total kinetic energy (CF valley) and for those with the small total kinetic
energy (EF valley).

Contrary to the estimates of Ref. \cite{Cwi89}, the results obtained by Moeller
et al. \cite{Moe87,Moe89,Moe92} indicate the existence of the second barrier on
the EF path. The authors have found even a third {\it switch-back} path going
from the SF valley via the second saddle to the EF valley. These calculations
are based on the finite-range liquid drop model and the  folded-Yukawa single
particle potential. Additionally, in Refs.  \cite{Moe87,Moe89} a smaller mass
parameter is postulated along the CF  path in order to obtain the comparable
spontaneous fission half-life time for both fission modes of $^{258}$Fm. In
Refs. \cite{Moe00,Moe01}  Moeller et al.  tried to find a solution to this
problem by making calculations on a 5-dimensional space using a finite range
liquid droplet model. They have also found two paths leading to fission, but
the low energy mode was presented as an asymmetric one, with mass asymmetry
$M_H/M_L=152.2/105.8$, which does not agree with the experimental data. It was
shown in \cite{Hul86,Hul89} that in $^{258}$Fm both modes of fission  only lead
to a symmetric split of the nucleus.

There are also some estimates of fission barriers made within the constrained
Hartree-Fock-Bogoliubov  (HFB) approximation with Gogny or Skyrme effective
interactions both at zero spin 
\cite{Bar82,Bra85,Ber84,Ber89,Ber96,Ber92,Cwi96,Rut97,Ben98} 
and at high spin
\cite{Egi00}. Most of these calculations where made for nuclei with $Z < 100$,
where only the traditional fission path (EF) appears in the experimental data.
In \cite{Ben98} calculations for $Z>100$ showing the compact fission path (CF)
where carried out for both non-relativistic (Skyrme interaction with the SkI4
parameters) and relativistic (in the context of the relativistic mean field
with the PL-40 parameterization) frameworks.

Recapitulating we can say that nuclei in the Fm region represent a very 
wide variety of fission types. The low mass isotopes fission with an
asymmetric mass distribution of the  fragments. The spontaneous fission
half-life ($T_{sf}$) increases from about 0.8~ms for $^{242}$Fm up to 126 y for
$^{252}$Fm and then falls down again to 0.36~ms for $^{258}$Fm in order to grow
again by one order of magnitude for $^{260}$Fm. The situation changes
especially dramatically when one goes from $^{256}$Fm to $^{258}$Fm. The
fission half-life decreases by 7~orders of magnitude and reaches 0.36~ms and,
as was described above, a very narrow symmetric mass  distribution appears in
the fission yield. Moreover, a similar behavior characterizes heavier nuclei
in the neighborhood of $^{258}$Fm. Up to now this rapid change of systematics
of $T_{sf}$ has not been well described theoretically \cite{Moe01,Sta99}. 

The aim of the present investigation is to look at the form of the fission
barriers obtained with Gogny forces for the nucleus $^{256}$Fm and its 
neighbors. In particular our aim is to answer the following questions: 
\begin{itemize}
\item[-] do there exist two fission valleys for nuclei in this region?
\item[-] is it possible to reproduce the experimental mass distributions of 
         the fission fragments?
\item[-] is it possible to explain the rapid change of the measured fission 
         half-life between $^{256}$Fm and $^{258}$Fm? 
\end{itemize}

In order to answer the above questions we have performed 
constrained Hartree-Fock-Bogoliubov calculations
as described in Sect.~2, where we give a brief outline of the
theoretical model as well as the description of the forces and the
configuration space. In Sect.~3 we discuss the results obtained.  Sect.~4
contains a summary and some concluding remarks.


\section{Theoretical model}

The Gogny density-dependent effective nucleon-nucleon force is taken in the
following form~\cite{Dec80}:
\begin{eqnarray}
V_{12}& = & \sum_{i=1}^2  \ (W_i + B_i\hat P_\sigma - H_i\hat P_\tau
-M_i\hat P_{\sigma}\hat P_{\tau}) \ e^{-\frac {\displaystyle
(\vec r_1- \vec r_2)^{2}} {\displaystyle\mu_i^{2}}}
\nonumber \\
& + & i \ W_{\rm LS}{\displaystyle \
(\overleftarrow{\nabla_1 - \nabla_2})\times\delta(\vec r_1 - \vec
r_2)(\overrightarrow{\nabla_1 - \nabla_2})} \cdot
 (\vec \sigma_1 + \vec \sigma_2)
\label{force} \\
& + & t_0 \ (1+x_0 \hat P_\sigma) \
\delta{\displaystyle (\vec r_1 - \vec r_2)}
\left[\rho(\frac{\displaystyle  \vec r_1 + \vec r_2 }
                {\displaystyle 2}) \right]^\gamma
+ V_{\rm Coul} \,\,, \nonumber
\end{eqnarray}
which contains a central finite range interaction, a zero-range spin-orbit term
and a zero-range density dependent interaction, respectively. The Coulomb 
interaction has to be added in the case of protons. The central interaction is
a sum of two Gaussian with widths $\mu_1$ and $\mu_2$. $\hat P\sigma$ and $\hat
P\tau$ denote the spin and isospin exchange operators respectively, and $\rho$
is the total density. 

We use the D1S \cite{Ber84,Ber91} parameterization of the Gogny interaction. The D1S
parameters were adjusted \cite{Ber84} to give a better surface energy term (crucial for
a proper description of the fission phenomenon) and their numerical values
are given by:
\begin{eqnarray}
W_1 = -1720.30 \;\ {\rm MeV} & \, \, \;\   W_2 = \,\, 103.639 \;
\ {\rm MeV} \nonumber  \\
B_1 =\;\ \, 1300.00 \;\ {\rm MeV} & \;\  \;\  B_2 = -163.483 \;\ {\rm MeV}
\nonumber \\
H_1 = -1813.53 \;\ {\rm MeV} & \, \, \, \;\  H_2 = \, \, \, \, \,  162.812 \;
\ {\rm MeV}   \nonumber \\
M_1=\;\ \, 1397.60 \;\ {\rm MeV} & \, \, \,   \;\ M_2=-223.934 \;\ {\rm MeV}
\label{D1S} \\
\mu_1= \;\, \qquad 0.7 \,\;\ \;\ \, {\rm fm} &  \;\   \mu_2= \qquad   1.2 \;
\,\,\,\,  {\rm fm} \nonumber \\
t_0=1390.6 \;\ {\rm MeV}\, {\rm fm}^{3(1+\gamma)} &  \! \! \! \! \! \! \! \! \!
 \! \!  \! \! \! \!   x_0= \qquad 1 \nonumber \\
 \gamma= \qquad 1/3 \quad \ \qquad  & \ \ \ W_{LS}= \quad 130 \ {\rm MeV} \,
 {\rm fm}^5 \nonumber
\end{eqnarray} 
The choice of the Gogny force with the D1S parameterization is based on the fact
that whenever this interaction has been used to describe low energy nuclear
structure phenomena an, at least, reasonable agreement with experiment has always 
been obtained. This degree of agreement has been obtained both for calculations 
at the mean field level and beyond.
\cite{Dec80,Ber84,Ber89,Ber96,Ber92,Egi00,Ber91,gir83,egi89,gir89,egi93,bla95,val96,gar98,Ang01,Rod02,Rod01}

In the microscopic HFB calculations we have used the computer code of
\cite{Egi97} where special attention was paid to an accurate computation of the
matrix elements of the Gogny interaction for very big bases like the ones used
in this paper.
The self-consistent equations have been
solved  by expanding  the  quasiparticle creation and annihilation
operators on finite bases of axially symmetric deformed harmonic oscillator
(HO) eigenfunctions.  The size of the bases used depend upon two parameters,
$N_0$ and $q$, which are  related to the allowed range of the HO quantum
numbers trough the relation
$$
\frac{1}{q} n_z + (2n_\perp + |m|) \le N_0.
$$
Along the perpendicular direction we take $N_0$ shells, (i.e. $2n_\perp +
|m|=0,\ldots,N_0$) and along the $z$ direction we include up to $qN_0$ shells
depending on the value of $2n_\perp +|m|$. In the present study we have used
$q=1.5$, a value which is suited for the elongated shapes along the $z$
direction typical of the fission process, and $N_0=13$, 15, and 17.  The reason
to use different values of $N_0$ is to study the convergence of our results
with the basis size. Another parameters characterizing the HO bases are the
oscillator lengths $b_\perp$ and $b_z$. These two quantities have been
determined, for each calculated wave function, as to minimize the HFB energy for
the $N_0=13$ basis. The same values of $b_\perp$ and $b_z$ are then used in
subsequent calculations with $N_0=15$ and 17 (see below for a discussion of the
convergence).

To study triaxiality effects in the first fission barrier we have also 
carried out calculations where
the axial symmetry requirement was released but the left-right symmetry was
imposed. As these calculations are much more time consuming than the axially
symmetric ones we had to restrict them to the $N_0=13$ case but, as it will be
discussed later, this is not a limitation in the region of interest.

In order to study the different paths to fission we have used in our
calculations  the following constraints: the axial quadrupole ($Q_2$), octupole
($Q_3$) and hexadecapole ($Q_4$) moments as well as the number of nucleons in
the neck region ($Q_N$). The corresponding operators are given by:
\begin{equation}
\hat Q_\lambda=r^\lambda P_\lambda (\cos (\theta)) ~~~~~~~~{\rm and}
~~~~~~~~\hat Q_N=\exp\left({-z^2\over a_N^2}\right) \,\,,
\end{equation}
with $a_N$=1~fm.

In the minimization process 
neither the two body kinetic
energy correction nor the Coulomb and spin orbit pairing energies have been taken
into account. Additionally, the Coulomb
exchange energy has been treated in the Slater approximation \cite{tit74,Ang01}. The reasons  are
the following: First, the calculation of the Coulomb exchange and pairing
energies is extremely time consuming \cite{Ang01} and its inclusion would
prevent the large scale calculations presented in this paper. From \cite{Ang01}
we know that Coulomb pairing can be very important for collective masses but
has little influence in the energy landscape. On the other hand, the Slater
approximation to the Coulomb exchange energy works fairly well in all the cases
(spherical or deformed nuclei) and is an affordable and reliable approximation.
Concerning the spin-orbit pairing its contribution to the pairing field  is
very small, specially at zero spin, and can be safely neglected. Finally, the
two body kinetic energy correction (2b-KEC) is not included in the variation
process because, for heavy nuclei, it remains almost constant for most of the
physical configurations. As this  term was included in the fitting of the
force, we have to include its contribution at the end of the calculation in
order to obtain reasonable binding energies.

We have also subtracted from the HFB energy the  rotational energy (REC) 
corrections stemming from the restoration of the rotational symmetry. This
correction has a considerable influence on the energy landscape (and therefore
on the height of the fission barriers) as is somehow proportional to the degree
of symmetry breaking and therefore proportional to the quadrupole moment. A
full calculation of the REC would imply an angular momentum projection
\cite{Rod02,Rod01} which is only feasible for light nuclei. In order to
estimate the REC we have followed the usual recipe \cite{Ring80} of
subtracting to the HFB energy the quantity  $ \langle \Delta
\vec{J}^{2}\rangle /(2{\mathcal{J}}_{Y}) $, where  $ \langle \Delta
\vec{J}^{2}\rangle $ is the fluctuation  associated with the angular momentum 
operators in the HFB wave function and $ {\mathcal{J}}_{Y} $  is the Yoccoz moment of
inertia \cite{Ring80}. This moment of inertia has been computed using the
``cranking'' approximation in which the full linear  response matrix appearing
in its expression is replaced by the zero order  approximation. The effect of
the ``cranking approximation" in the Yoccoz moment of inertia was analyzed with
the Gogny interaction for heavy nuclei in \cite{Egi00} by comparing it  with
the one extracted from an angular momentum projected calculation (see also
\cite{Rod01} for a comparison in light nuclei). The conclusion is that the
exact REC is a factor 0.7 smaller than the one computed with the ``cranking"
approximation to the Yoccoz moment of inertia for strongly deformed
configurations (a similar behavior has been observed for the Thouless-Valatin
moment of inertia in \cite{Gir92}). We have taken this phenomenological factor
into account in our calculation of the REC.

In the last section we analyze the spontaneous fission half life of several
Fm isotopes. The analysis was carried out in the standard WKB framework where
$T_{\rm sf}$ is given (in seconds) by 
\begin{equation}
T_{\rm sf} = 2.86 \cdot 10^{-21} (1+\exp (2S)) \,\,.
\end{equation}
In this expression $S$ is the action along the $Q_2$ constrained path which is
given by 
\begin{equation}
S = \int_a^b dQ_2 \sqrt{2B(Q_2)(V(Q_2)-E_0)}\,\,.
\end{equation}
For the collective quadrupole inertia $B(Q_2)$ we have used the ATDHFB 
expression computed again in the ``cranking" approximation and given by
\cite{Gia80} 
\begin{equation}
 B_{ATDHFB}(Q_2) = \frac{M_{-3}   (Q_2)}		    
               {M_{-1}^2 (Q_2)} \,\,,
\end{equation}
with 
\begin{equation}
M_{-n} (Q_2) = \sum_{\mu \nu} 
               \frac{ |Q^{20}_{\mu \nu}|^2 }
	            {   (E_\mu + E_\nu)^n  } \,\,.
\end{equation}
Here $Q^{20}_{\mu \nu}$ is the 20 component of the quadrupole operator
$\hat{Q}_2$ in the
quasiparticle representation \cite{Ring80} and $E_\mu$ are the quasiparticle
energies obtained in the solution of the HFB equation.

In the expression for the action the collective potential $V(Q_2)$ is given 
by the HFB energy (with the 2b-KEC and REC corrections) minus the zero point 
energy (ZPE) correction $\epsilon_0 (Q_2)$ associated with the quadrupole 
motion. This ZPE correction is given by 
\begin{equation}
\epsilon_0 (Q_2) = \frac{1}{2} G(Q_2) B^{-1}_{ATDHFB} (Q_2) \,\,,
\end{equation}
where 
\begin{equation}
 G(Q_2) = \frac{M_{-2}   (Q_2)}		    
               {2M_{-1}^2 (Q_2)} \,\,.
\end{equation}
Finally, in the expression for the action and additional parameter $E_0$ is 
introduced. This parameter can be taken as the HFB energy of the (metastable)
ground state. However, it is argued that in a quantal treatment of the problem
the ground state energy is given by the HFB energy plus the zero point energy
associated to the collective motion. To account for this fact,  the usual
recipe is to add an estimation of the zero point energy to the HFB energy in
order to obtain $E_0$. In our calculations we have taken a zero point energy of
0.5 MeV for all the isotopes considered. 


\subsection{Convergence of the calculations}

In our calculations  bases with $N_0=13, 15$ or 17  were used
in order to check  the convergence of the results. As an example of these tests
we display in Fig.~1 the HFB energies corresponding to the CF path of 
$^{256}$Fm  as a function of $Q_2$ for  different values of $N_0$. A comparison
of the $N_0=13$ and the $N_0=15$ results show that $N_0=13$ is enough in the
region around the first barrier but this is not the case in the region of the
superdeformed minimum, around $Q_2=100$~b,  where a 4 MeV shift is observed in
going from $N_0=13$ to $N_0=15$.

Using $N_0=17$  we  obtain rather stable results as compared with the $N_0=15$
calculations. That is, for most of the $Q_2$ range the difference between the
$N_0=17$ and $N_0=15$ energies is almost independent for $Q_2$. The difference 
becomes visible only for $Q_2 > 200$ b, but this region is irrelevant to our
investigation as it corresponds to solutions with well separated fragments. 
This behavior is typical for all paths and nuclei presented in
this paper and from the above considerations, one can conclude that $N_0=15$ is
sufficient for the planned HFB calculations. We have also checked that this
fast convergence  with the basis size is a consequence of the optimization of
the oscillator lengths carried out for each quadrupole deformation. The
oscillator length parameters were chosen as to minimize the HFB energies in the
$N_0=13$ calculations.


\subsection{Two body kinetic energy and rotational energy corrections}

The influence of the two-body kinetic energy (2b-KEC) and the rotational energy
(REC) corrections on the binding energy and the fission barrier of $^{256}$Fm
is shown in Fig.~2. It is seen in the figure that the 2b-KEC shifts up the
binding energy by around 13~MeV with respect the HFB estimate, while the REC is
negative and its magnitude grows from 0 for the spherical configuration  to
about 5~MeV in the region of the second barrier (see inset). The plateau
observed in the REC starting in $Q_2=140$b is due to the fact that from there on
the solution correspond to two separated spherical fragments. In this case both
the moment of inertia and the fluctuation of the angular momentum operators
are proportional to  the square of the distance between the fragments.
As a consequence of the behaviour of  the REC as a function of the quadrupole
moment, its inclusion
decreases the second barrier by a few MeV and that has an important influence
on the systematics of the spontaneous fission life-times of heavy Fm isotopes
as it will be seen in the next section. All potential energy surfaces (PES)
presented in this paper contain both the 2b-KEC and REC corrections described
above.


\section{Results}

All nuclei considered here have similar barrier shapes exhibiting two humps.
The ground state minimum is at approximately $Q_2$=15~b, which corresponds to
the deformation parameter $\beta_2 \approx 0.25$. The first fission barrier is,
in our axially symmetric calculations, about 10~MeV high but it is decreased by a few MeV when 
triaxial shapes are included in the analysis what is in agreement with results
of other authors (consult e.g. Refs. \cite{Bar81,Ben98}). The lowering of the
fission barrier due to triaxiality comes together with an increase of the
collective mass and therefore the effect of triaxiality on the fission
half-lives is rather small. This is in agreement with the results
of  Ref. \cite{Bar81} where it was found that the least action trajectory, or in other words
the dynamical path to fission, leads only through the axially symmetric shapes
of fissioning nucleus. Finally, a superdeformed minimum  at an energy similar to the 
ground state energy appears at $Q_2 \approx 50$~b. It is separated from the
scission point by a small second barrier that, as we will see in the next
section, plays a fundamental role in the fission half-lives.


\subsection{Nuclear Properties along the fission paths}

  In this section the potential energy $E$, the octupole $Q_3$ and hexadecapole
$Q_4$ moments as well as the neck  parameter $Q_N$  are investigated  along the
fission paths  for the  nuclei $^{256,258}$Fm,  $^{258}$No and $^{260}$Rf.

\subsubsection{$^{256}${\rm Fm}.}

In  Fig.~3, the results of the calculations for the nucleus $^{256}$Fm are
plotted  for the CF  path (solid lines) and the EF path (dashed lines). 
In panel (a) we show the potential energy surfaces as a function of the
quadrupole moment for both paths. Along with the energy curve we have also
plotted the real shape of the nucleus for relevant values of the quadrupole
moment. The reduction of the first barrier  by
approximately 4 MeV due to the triaxial degrees of freedom ($\gamma$ is
typically in the range between 0 and 8 degrees) is marked by the dotted
line. We observe that after tunnelling
through the first barrier the nucleus goes into the  superdeformed region. 
In fact there are two superdeformed minima, one  at $Q_2=$50 and another at 70~b
separated by a tiny barrier. The deeper minimum at  $Q_2=70$~b is situated
2~MeV below the ground state energy. However, it is only separated from the
scission point at $Q_2=130$ by a barrier which is only 2 MeV high and therefore
it is rather unlikely that this superdeformed (ground state) minimum can live
long enought as to be considered a metastable state.
The fission products corresponding to this path are
identical and spherical, in fact, the fragments are two spherical $^{128}$Sn 
nuclei. Such type of fission path (solid line in Fig.~3) was called in Ref.
\cite{Cwi89} the compact fission (CF) path. It was also shown in \cite{Cwi89}
that the octupole moment along such a path is equal to zero what is in line
with our results, see panel (b) of Fig.~3. After passing the scission point the
potential energy (the Coulomb energy in fact) of this fissioning system
decreases rapidly with growing quadrupole moment. The other path, called
elongated fission (EF) path,  begins at $Q_2$=70~b. This path plays a crucial
role in the fission process of this nucleus. It corresponds to the reflection
asymmetric shapes with $Q_3\ne0$ as one can observe in panel (b) of Fig.~3.
Both fission paths differ also significantly in the hexadecapole moments, panel
(c), and in the number of nucleons in the neck region, panel (d).  For 
quadrupole moments larger than 120~eb one finds  that the EF path has a 
gentler slope than the CF path.  

In order to understand the shapes of the EF path for this nucleus and the
heavier isotopes considered we have plotted in panel (a) of Fig. 4
the shape of $^{256}$Fm  at the 
deformation  $Q_2=200$~b. On the left-right asymmetric shape
distribution of the fissioning nucleus one can distinguish two fragments
connected by a neck. One of the fragments is close to a sphere and the other
one has a rather large quadrupole deformation. In order to study the mass
contents of both fragments we have plotted in panel (b) the quantity 
$$
N(z)= 2\pi \int_{-\infty}^{z} dz' \int_0^\infty dr_\perp \rho(r_\perp z')
$$
for both protons and neutrons. The number of particles corresponding to the
magic numbers 50 and 82 are marked in the panel (b) with horizontal dotted
lines. From this plot we learn that  both fragments have roughly the same mass
and they correspond to Sn isotopes close to the doubly magic $^{132}$Sn. The
fact that an strongly left-right asymmetric mass distribution leads to two
fragments with roughly the same mass is a remarkable result that will be
commented later. Finally, no significant lowering of the density is observed in
the neck region (panel (c) of Fig. 4). 

The transition from the CF to
the EF path take  place first at $Q_2=90$~b because, for smaller quadrupole
moments, the paths are separated by a 5~MeV high ridge as can be seen in Fig.~5,
where cross sections of the potential energy surface for various quadrupole
moments are plotted as a function of the neck parameter $Q_N$. When the barrier
between both valleys disappears the nucleus continues fission along the EF path.
Such a behavior, referred to by other authors a "switchback path" \cite{Moe92}
seems to be energetically most preferable. From $Q_2$=100~b up to 130~b the
minimum corresponding to the CF valley  becomes a shoulder as it is seen in
Fig.~5. This means that $^{256}$Fm can not continue fissioning along this mode
and will proceed through the EF path explaining the low kinetic energy
distribution of the fission fragments of this nucleus. 
The minimum corresponding to the CF valley appears again at $Q_2$=140~b but at
such a large quadrupole moments both fragments are already separated. At
$Q_2$=140~b the fission valleys CF and EF are separated by a 4~MeV high
barrier.

The fragments which are created in the EF process of $^{256}$Fm have
different deformations but nearly equal masses as seen in Fig. 4. 
This is inconsistent with the experimental mass distribution 
which shows a mass asymmetry $A_H/A_L\sim
141/112$ \cite{Van73}. Our static calculations are based only on the PES of the
fissioning nucleus and it seems that dynamical effects could play a certain
role in the fission of $^{256}$Fm. It could also happens, but this is
rather less probable, that we are not able to find such a mass asymmetric path
in our calculation. Apart from the two valleys described above a few others
paths were found. All of them are localized much above the CF and EF paths so
it is rather improbable that the fissioning nucleus will follow one of them.


\subsubsection{$^{258}${\rm Fm.}}

It was found experimentally in Ref. \cite{Hul86,Hul89} that the nucleus
$^{258}$Fm exhibits bimodal fission. Both modes have similar abundance and
symmetric mass distribution. In panel (a) of Fig.~6 the CF (solid line) and EF
(dashed line) fission barriers are shown. The  octupole, hexadecapole moments
and the number of nucleons in the neck region $Q_N$ corresponding to both paths are
presented in panels (b-d) of Fig.~6. Generally speaking the picture is very
similar to the one of $^{256}$Fm, but there are some important differences
which cause changes in the fission yield  of $^{258}$Fm. 

The first distinction between this nucleus and $^{256}$Fm is the fact that the
second hump of the fission barrier on the CF path (at $Q_2$=120~b) has
practically disappeared and rises only 0.5 MeV above the superdeformed minimum.
Additionally the top of the second barrier is placed a few MeV below the ground
state energy. 

The second difference with respect to $^{256}$Fm is the relation between the
two paths leading to fission. As it is seen in Fig.~7, the minima corresponding
to the CF and EF coexist along the fission path, i.e., the ridge between them
does  not disappear along the whole way to fission. Its height always exceeds
1.5 MeV. It means that the nucleus could fission via the CF valley or change
the path and proceed with the elongated fission (EF) fragment path starting
from $Q_2\approx$ 90~b. At higher quadrupole moments
the transfer from the CF to the EF path is also possible but it is less
probable as the ridge between the paths in the region of the second hump rises
up to 3~MeV. Such a configuration of the EF and CF paths seems to ensure
(because we do  not account here for dynamical effects) that both modes are 
fed with similar intensity. This result is in agreement with conclusions of
Refs. \cite{Hul86,Hul89}, where comparable abundance of both modes was found.

We can identify the CF path with the high total kinetic energy (TKE) mode in
the fission of $^{258}$Fm and the EF path with the low TKE mode. In the CF path
the nucleus splits into two identical, spherical parts which are two $^{129}$Sn
nuclei. At the scission point, the distance between the centers of masses of
these spherical fission fragments is relatively small what gives a strong
Coulomb repulsion and in consequence a high mean value of the TKE of the
fragments. The fissioning nucleus passing through the second EF path has a
similar elongated shape as the one described for $^{256}$Fm in Fig. 4. The distance
between the mass centers of the two born fragments is much larger than the one
for the CF path. It causes a weaker Coulomb repulsion between fragments and in
consequence a smaller mean TKE of the fragments.

We have got also some arguments in favor of the hypothesis that in the fission
of $^{258}$Fm one deals with a kind of cluster fission \cite{Goe99}, both in
the CF and in the EF paths. Looking at the proton and neutron density
distribution along both valleys we have found that the nucleus $^{258}$Fm
splits into two parts with equal masses. Each fragment has around 50 protons
and 79 neutrons. The only difference is that one of the $^{129}$Sn fragments
born in
the EF path is highly deformed with $\beta_2=0.6$ whereas is spherical in the CF
path. As the different TKE of the two paths could be explained in terms of the
energy difference between the spherical and the superdeformed fragment, we have performed
additional  $Q_2$-constrained HFB calculations for a few Sn isotopes. The
results for $^{126-132}$Sn even--even isotopes are presented in Fig.~8. A
shoulder (or even flat minimum for $^{130}$Sn) is seen for all isotopes at
$Q_2$=10 to 14~b. This superdeformed second minimum for $^{130}$Sn corresponds
to $Q_2$=12~b (or $\beta_2=0.6$) and it is located  around 23~MeV above the
ground state. This superdeformed state can be identified as the deformed
$^{129}$Sn fragment meaning that the 23 MeV  accumulated in the superdeformed 
state will be  taken away by post--fission neutrons or
$\gamma$--rays and will not be converted to kinetic energy of the fragments. 
Therefore, we expect that the TKE of the fragments for the EF path to be of the
order of 23 MeV smaller than for the CF path, in good agreement with experiment.


\subsubsection{$^{258}${\rm No and }$^{260}${\rm Rf}.}

Figs.~9 and 10  (Figs.~11 and 12) show the PES and its cross sections for the 
fissioning $^{258}$No  ($^{260}$Rf) nucleus.  In these nuclei, the second hump
of the potential barrier disappears completely. After tunneling through the
first barrier they fission directly.

In the nuclei $^{258}$No (Fig.~9) and $^{260}$Rf (Fig.~11), as in $^{258}$Fm, 
the  CF and EF paths are also found. The transition between both  
valleys is possible at $Q_2$=90~b, as it can be seen in Figs.~10 and 12, where
there is no ridge separating them.  Here, similarly to the case of $^{256}$Fm,
the minimum corresponding to the CF path turns into a shoulder at about
$Q_2$=100~b, but it appears again at  $Q_2$=120 to 130~b. It makes possible the
come-back to the CF path, although the probability for such a process is
relatively small. This effect explains the experimentally observed low
abundance of the high TKE mode: 5\% for $^{258}$No and even less for 
$^{260}$Rf. 

In configurations close to scission another third path corresponding to two
compact fragments (marked by the dotted lines in Figs.~9 and 11) will appear.
In contrast to the above considered Fm isotopes, the two fragments  have a
small asymmetry in mass with a rate of $M_H/M_L$=132.5/125.5 for  $^{258}$No
and $M_H/M_L$=136/124 for  $^{260}$Rf. In both cases the number of protons is
the same in both fragments. This asymmetry is in line with the experimental
results for these nuclei \cite{Hul86,Hul89}.

\subsection{Spontaneous fission half-lives}

For heavy Fm isotopes the spontaneous fission half-lives ($T_{sf}$) decrease
rapidly with mass. This fall off has up to now (see e.g. \cite{Sta99}) not
found a satisfactory explanation.  

The shape of the potential barrier is one of the most important factors which
determines the fission half-life of nuclei. The fission barriers for
$^{254}$Fm, $^{256}$Fm and $^{258}$Fm (for the CF path only) are plotted in
Fig.~13. All curves are shifted in order to get the ground state minimum in the
same position. One can see in Fig.~13 that the first hump  of the barrier is
practically the same for  all these nuclei. In fact, the almost 10~MeV high
fission barriers are reduced by a few MeV when including the effect of
triaxiality (see in Figs. 3 and 6). However, as already mentioned, it was shown
in Ref. \cite{Bar81} that the dynamical effects prevents the fissioning nucleus
to take axially non symmetric forms, so we have decided to perform the estimates
of $T_{sf}$ for the both (axial and nonaxial) cases.

The main difference between these three isotopes is in the location of the
second hump with respect to the ground state minimum. In $^{256}$Fm the  second
barrier is  at the ground state level  and in  $^{258}$Fm  a few MeV  lower
than that. In these nuclei only the first hump will influence the  fission
half-life time. The fission barrier of $^{254}$Fm has a completely different
shape. Although, the shift up of the second hump with respect to $^{256}$Fm is
not larger than the corresponding shift in $^{256}$Fm with respect to 
$^{258}$Fm, this small shift causes that the whole second barrier is now above
the ground state. This effect influences the theoretical estimates for fission
half-lives in a dramatical way. The half-life for $^{254}$Fm is in our
estimates 11 orders of magnitude larger than that for $^{256}$Fm when the
symmetric (CF) path  (solid line in Fig.~3) is taken into account (open symbols
in Fig.~14).  It becomes much shorter (full symbols in Fig.~14), and almost
equal to the experimental one, when the reduction of the fission barrier due to
the left-right asymmetry degrees of freedom is included. It has obviously to do
with  the fact that the size of the fission  barrier to be tunneled in
$^{254}$Fm reduces slightly if one switches to the EF valley at
$Q_2\approx$90~b. The estimates done assuming the axially symmetric form of
fissioning nucleus are marked in Fig. 14 by the circles, while these for the
nonaxial case are denoted by the triangles. It is seen that the inclusion of
the nonaxial degrees of freedom decreases the values of $T_{sf}$ obtained in
our one dimensional calculation (i.e. without the dynamical effects of Ref.
\cite{Bar81}) by about one order of magnitude.  

In Fig. 14 we compare our estimates for the spontaneous fission half-lives
($T_{sf}$) with the experimental data and also with the results of  dynamical
calculations on the basis of the Saxon--Woods potential made in
Ref.~\cite{Sta99}.  It is seen that we have qualitatively explained its
decrease for heavy Fm isotopes. It is due to the fact that  the second hump
goes below the ground state level in $^ {256-258}$Fm. In fact, we did not
reproduce well enough the half-life time for $^{256}$Fm which is in our
calculations  3 orders of magnitude shorter than the measured value and
therefore, the abrupt change in the systematics of the fission life-times
appears 2 mass units too early.


\section{Conclusions}

In this paper we have discussed some properties of the potential energy
surfaces of fissioning even-even nuclei in the $^{258}$Fm region. All discussed
nuclei exhibit two-hump fission barriers. There is always one of the fission
trajectories which leads to compact fission. Another path, leading to an
alternative mode corresponding  to a spherical and an elongated fission
fragment, has been found for all these nuclei. The shape of the second hump of
the potential barrier and the relation between the two fission paths in the
potential energy surface are crucial for the way in which fission occurs in
these nuclei. 

In $^{256}$Fm fission follows only the elongated fission path but we are
unable, in our static calculation, to reproduce properly the mass asymmetry of
the fission products of $^{256}$Fm. Similarly to the experimental 
situation the theoretical approach yields only a low TKE mode in the fission
yields was found theoretically. In contrast to this, the isotope $^{258}$Fm
may fission along the CF or EF valleys and bimodal fission take place there.
The observed experimental difference of the TKE between both modes is well
reproduced by our model. The mechanism of bimodal fission of $^{258}$Fm and the
heavier even-even nuclei has been described properly. 

We also explained the decrease of the half-life times for the heavy Fm
isotopes. In the case of $^{254}$Fm, the second hump on the fission barrier is
located above the ground state. In heavier isotopes it goes down by a few MeV 
below the ground state minimum and therefore does not give any contribution to
the half-life times of these nuclei.

Contrary to the majority of papers describing bimodal fission of $^{258}$Fm we
have found a strong left-right asymmetry in the shape of fissioning nucleus
which nevertheless corresponds to a symmetric mass split. This is a new
phenomenon which could be discovered only in the HFB type of calculations with
the Gogny or $\delta$--pairing forces which distinguish between orbitals in
different fragments. \\
~

\noindent
{\bf Acknowledgments:  }

Two of the authors (K. P. and M. W.) gratefully acknowledge the warm
hospitality extended to them by the Departamento de F\'{\i}sica Te\'orica  of
the Universidad Aut\' onoma de Madrid as well as grants from the Spanish
Interministerial Commission of Science and Technology, Ref.~SB99-BA6182184 (M.
W.) and of the Spanish Foreign Ministry (K. P.).  

This work is partly sponsored by DGICyT, Spain under Project PB97-0023 and the
Polish Committee of Scientific Research, grant No.~2P~03B~115~19.

The critical reading of the manuscript and the comments done by Johann Bartel
is also acknowledged. 


\pagebreak

\begin{figure}
\begin{center}
\leavevmode
\epsfig{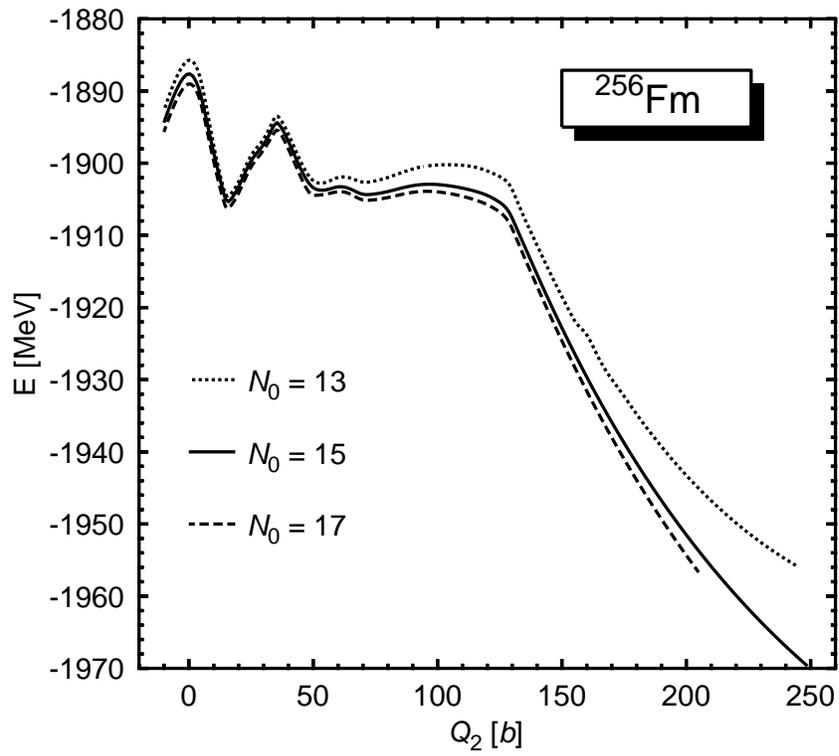}
\end{center}
\caption{The HFB energy obtained for $N_0$=13, 15 and 17 for the compact 
         fission (CF) path of $^{256}$Fm as a function of the quadrupole 
         moment $Q_{2}$.}
\label{fig1}
\end{figure}
                                                                                
\pagebreak

\begin{figure}
\begin{center}
\leavevmode
\epsfig{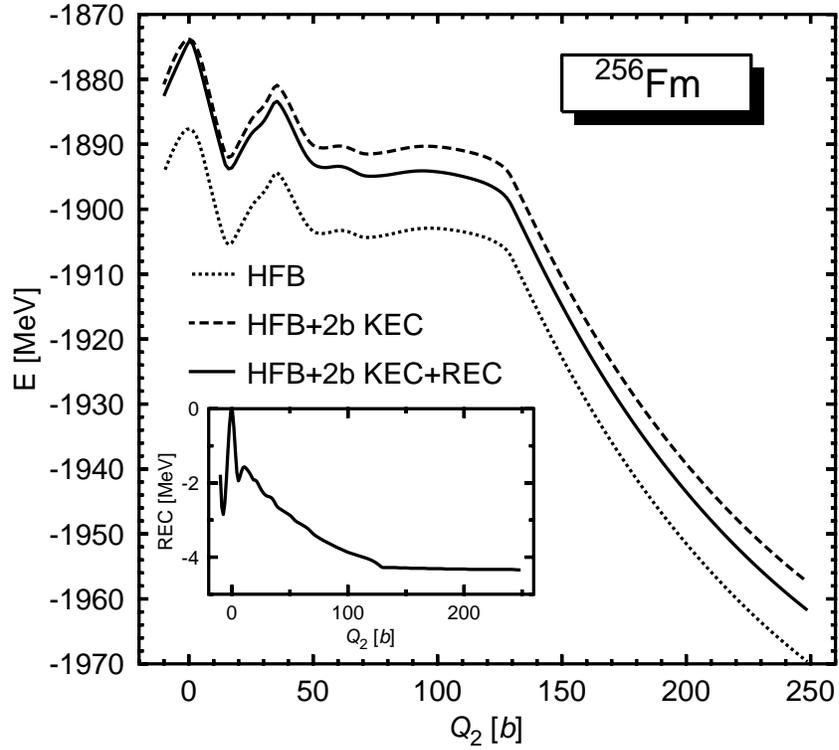}
\end{center}
\caption{The potential energy  surface obtained with the HFB
         calculations  (dotted line), the one taking into account the 
	 two body energy correction (dashed line) and the one including 
         also the rotational energy correction (full line) for the compact 
         fission  path of $^{256}$Fm as a function of the quadrupole moment 
         $Q_{2}$.} 
\label{fig2}
\end{figure}

\pagebreak

\begin{figure}
\begin{center}
\leavevmode
\epsfig{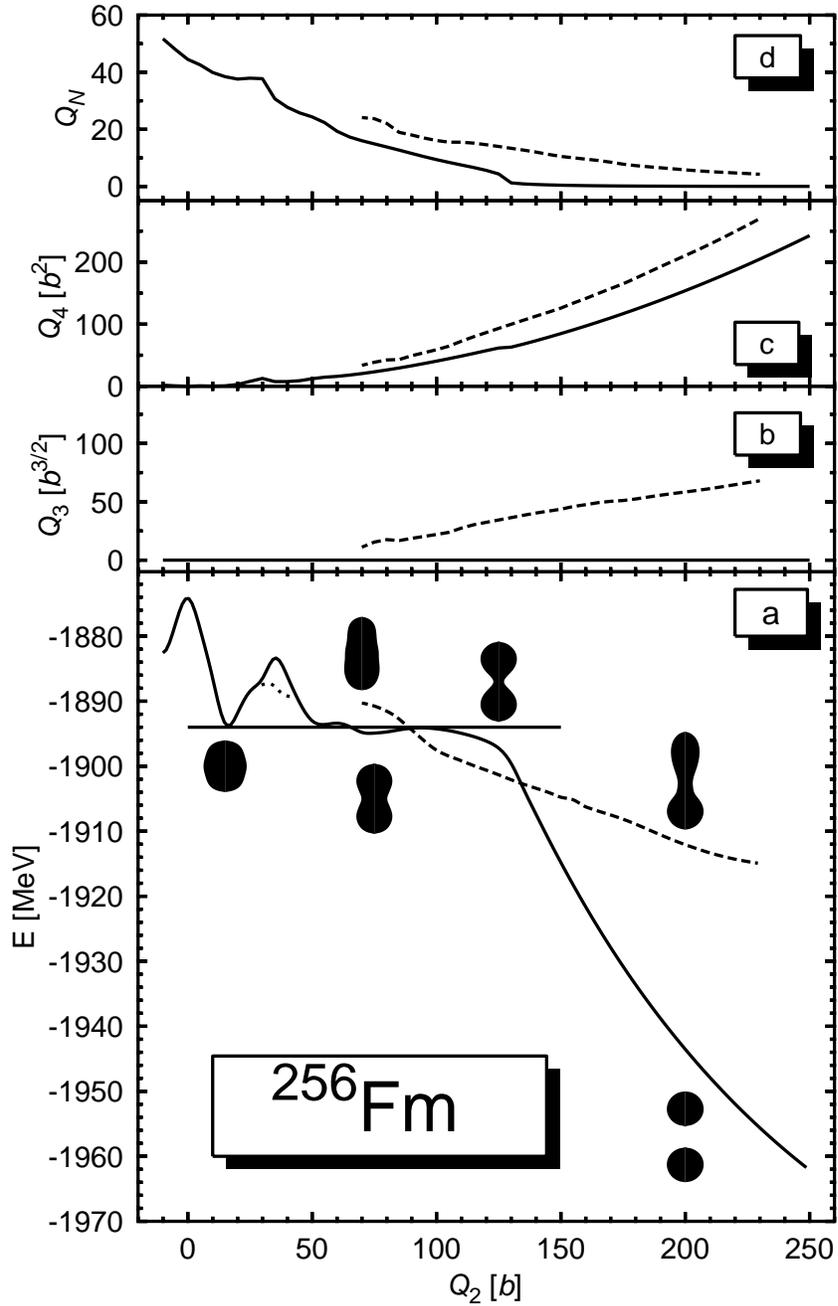}
\end{center}
\caption{Panel (a): The fission barrier of $^{256}$Fm as a function of the 
         quadrupole moment $Q_{2}$ for $N_0=15$. The solid line corresponds
         to the compact fission path (CF) and the dashed line to the elongated 
         one (EF). The dotted line shows the reduction of the first barrier
         due to nonaxial degrees of freedom. The shapes of the nucleus at 
         a density of $\rho_0=0.08 fm^ {-3}$ are depicted for several values of 
         $Q_2$ both for the CF and EF paths (note that the EF path leads to 
         octupole deformed shapes). Additionally, in panels (b), (c) and (d) 
         the octupole and hexadecapole moment as well as the neck parameter 
         $Q_N$ are respectively plotted.}
\label{fig3}
\end{figure}

\pagebreak

\begin{figure}
\begin{center}
\leavevmode
\epsfig{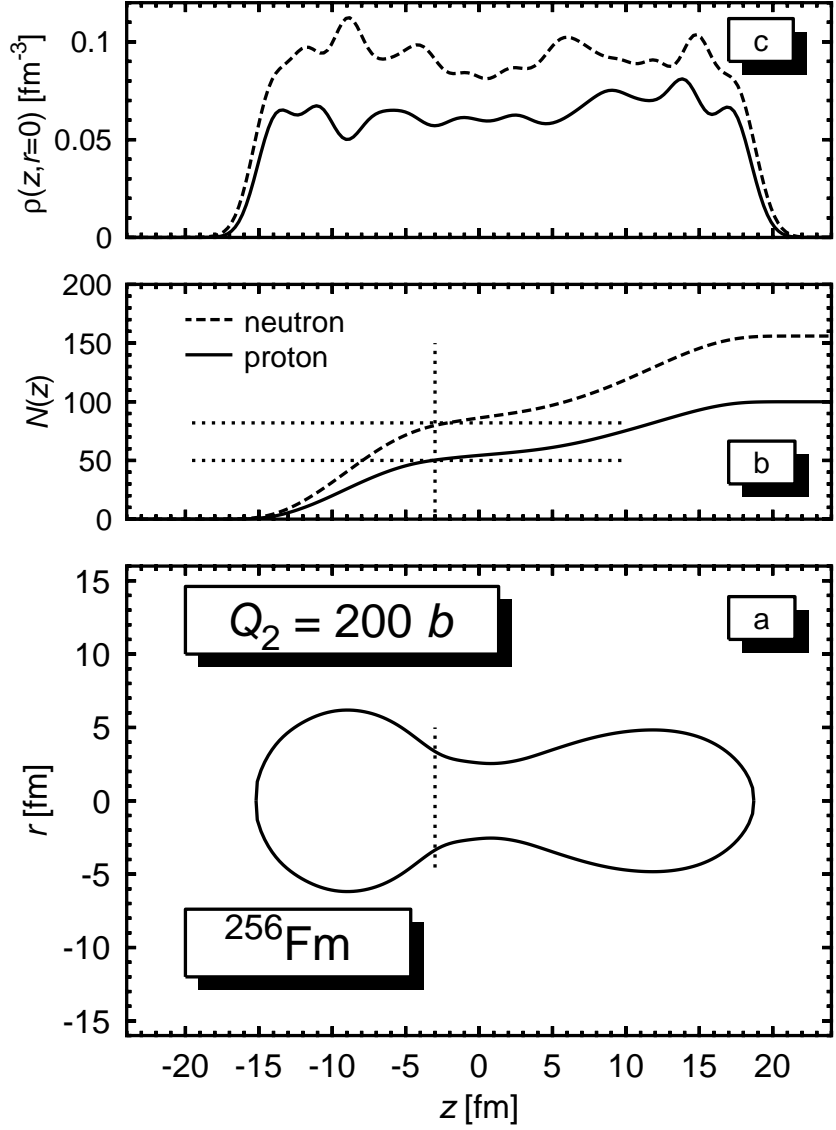}
\end{center}
\caption{The shape of $^{256}$Fm at deformation $Q_{2}=200$b on the elongated 
         path to fission as well as the number of particles (b) and the density 
         (c) of protons (solid line) and neutrons (dashed line) as a function 
         of $z$. The number of particles corresponding to the magic numbers
	 50 and 82 are marked in the panel (b) with horizontal dotted lines.}
\label{fig4}
\end{figure}

\pagebreak

\begin{figure}
\begin{center}
\leavevmode
\epsfig{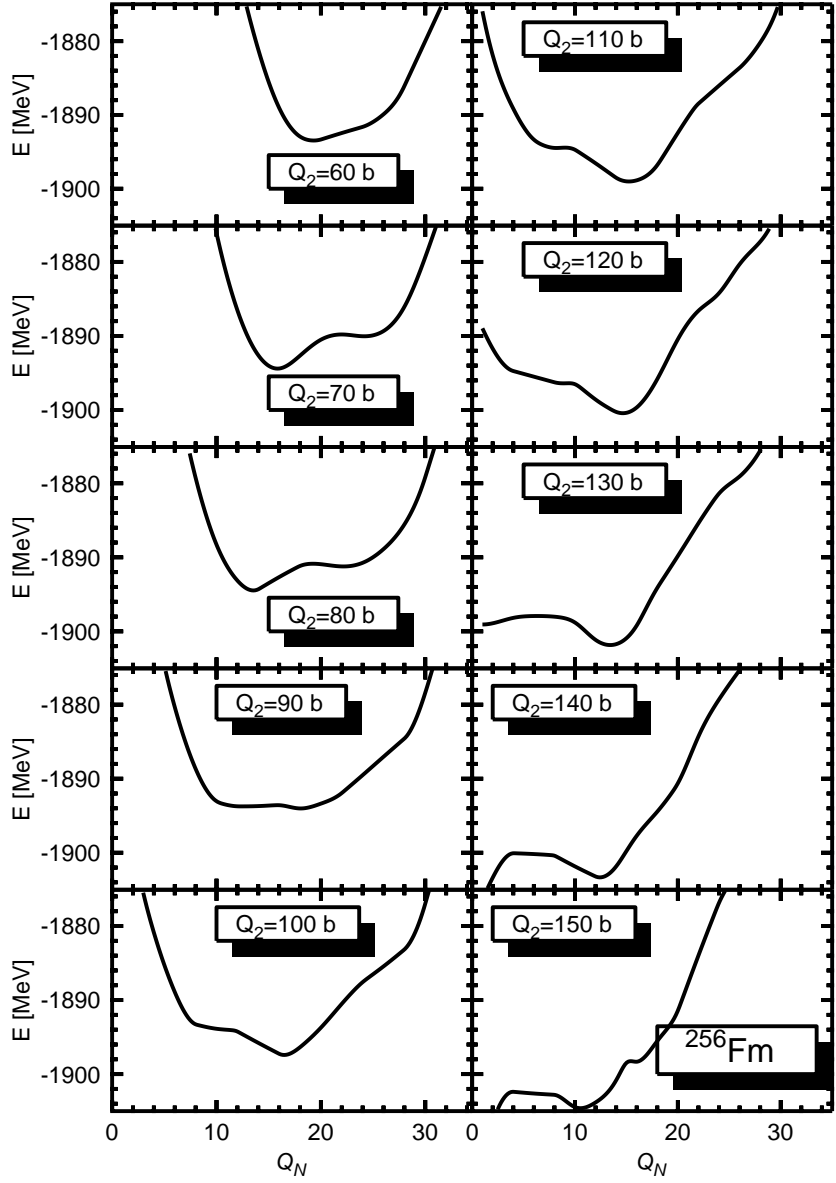}
\end{center}
\caption{The cross section of the potential energy surface of $^{256}$Fm for
         different values of $Q_{2}$ as a function of the neck parameter $Q_N$.}
\label{fig5}
\end{figure}

\pagebreak

\begin{figure}
\begin{center}
\leavevmode
\epsfig{file=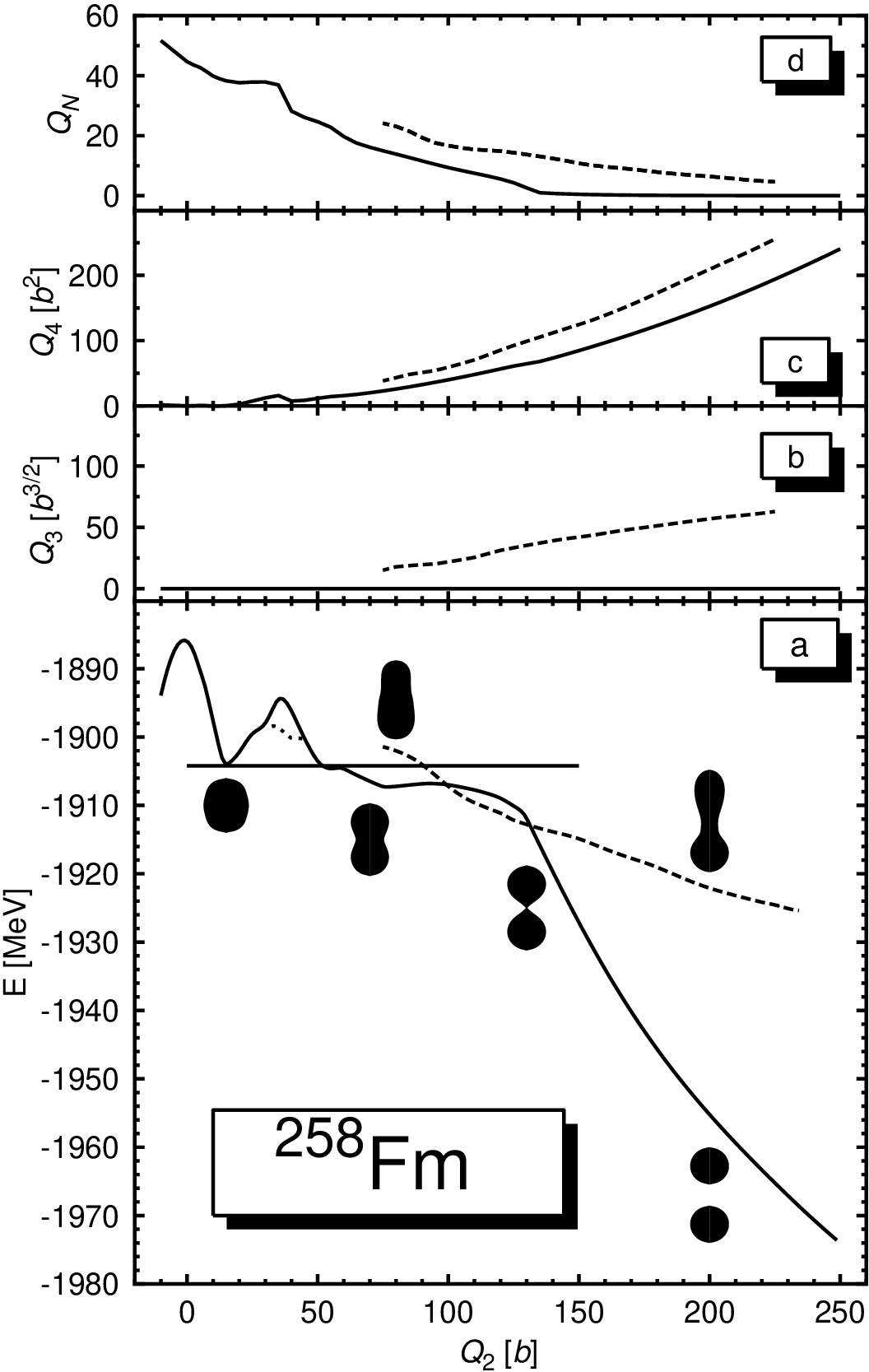, width=11cm, angle=00}
\end{center}
\caption{The same as in Fig.~3, but for the  $^{258}$Fm nucleus. }
\label{fig6}
\end{figure}

\pagebreak

\begin{figure}
\begin{center}
\leavevmode
\epsfig{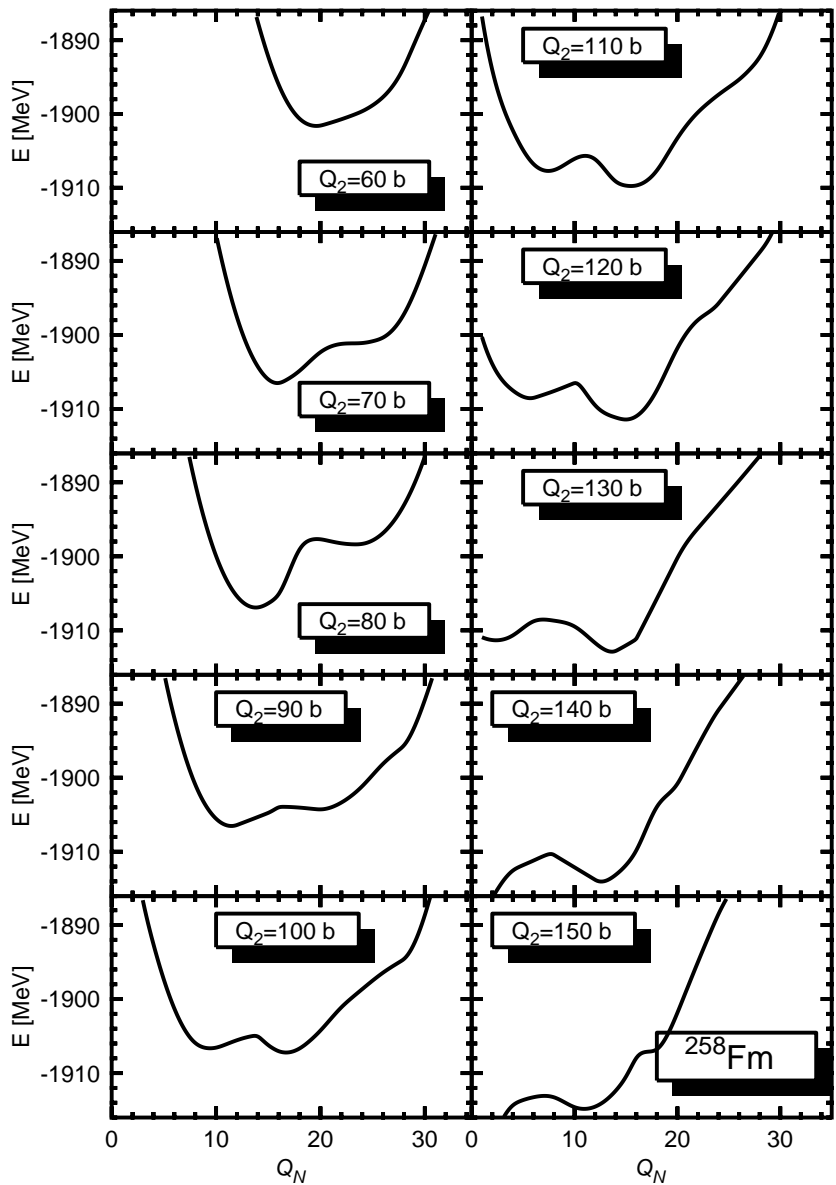}
\end{center}
\caption{The same as in Fig.~5, but for the  $^{258}$Fm nucleus.}
\label{fig7}
\end{figure}

\pagebreak

\begin{figure}
\begin{center}
\leavevmode
\epsfig{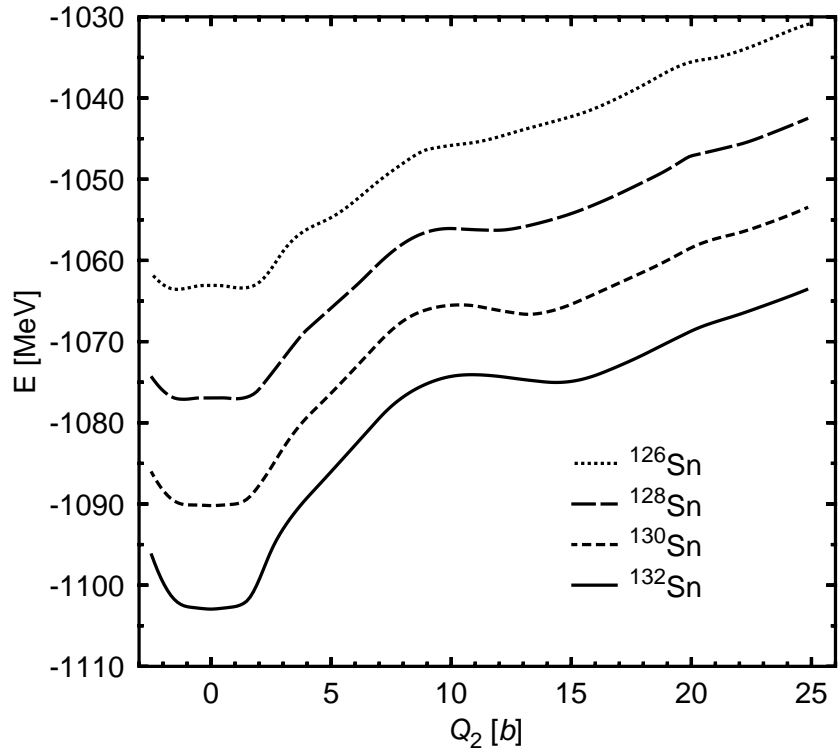}
\end{center}
\caption{The potential energy surfaces for some Sn isotopes as a function 
         of the quadrupole moment $Q_2$. Both the 2b-KEC and REC are included
	 in the curves}
\label{fig8}
\end{figure}

\pagebreak

\begin{figure}
\begin{center}
\leavevmode
\epsfig{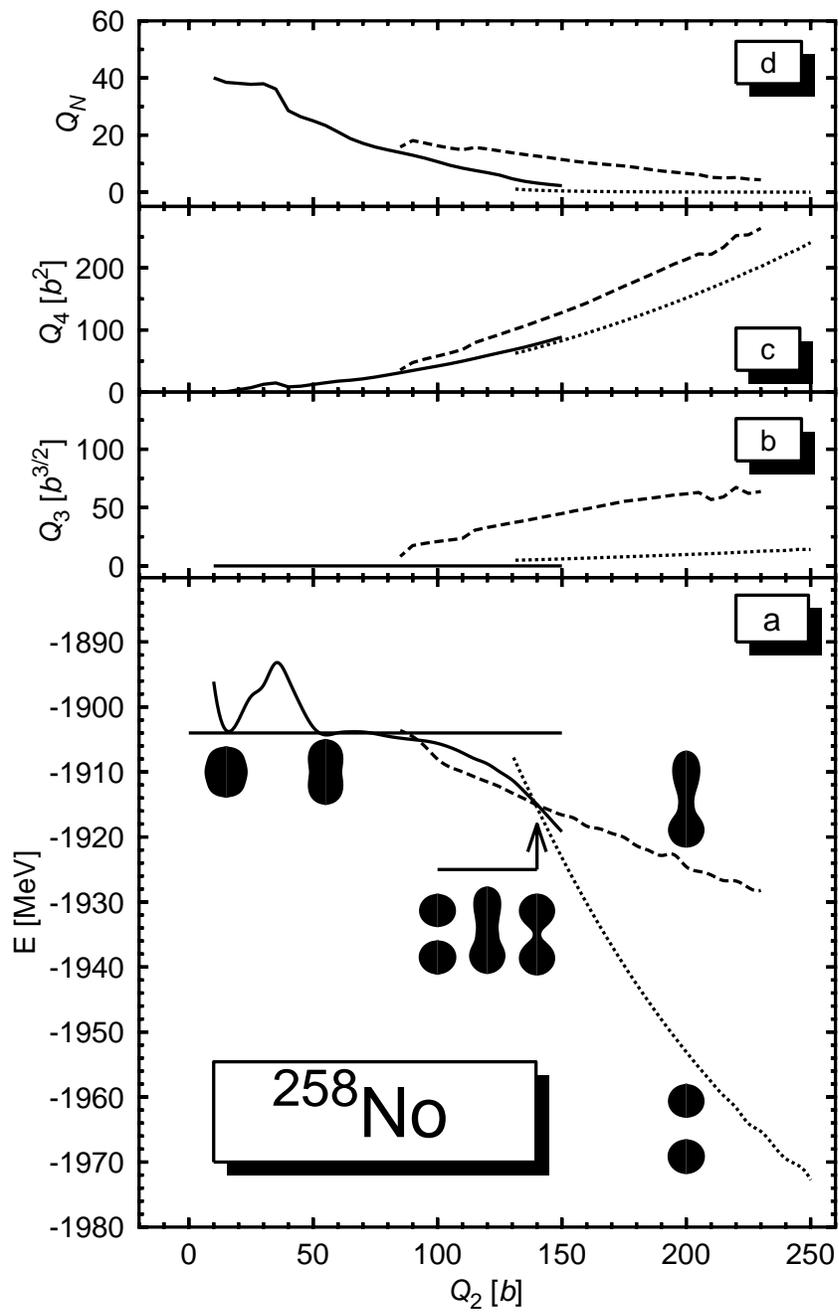}
\end{center}
\caption{The same as in Fig.~3, but for the  $^{258}$No nucleus. The dotted 
         line corresponds to the two-body solution. }
\label{fig9}
\end{figure}

\pagebreak

\begin{figure}
\begin{center}
\leavevmode
\epsfig{file=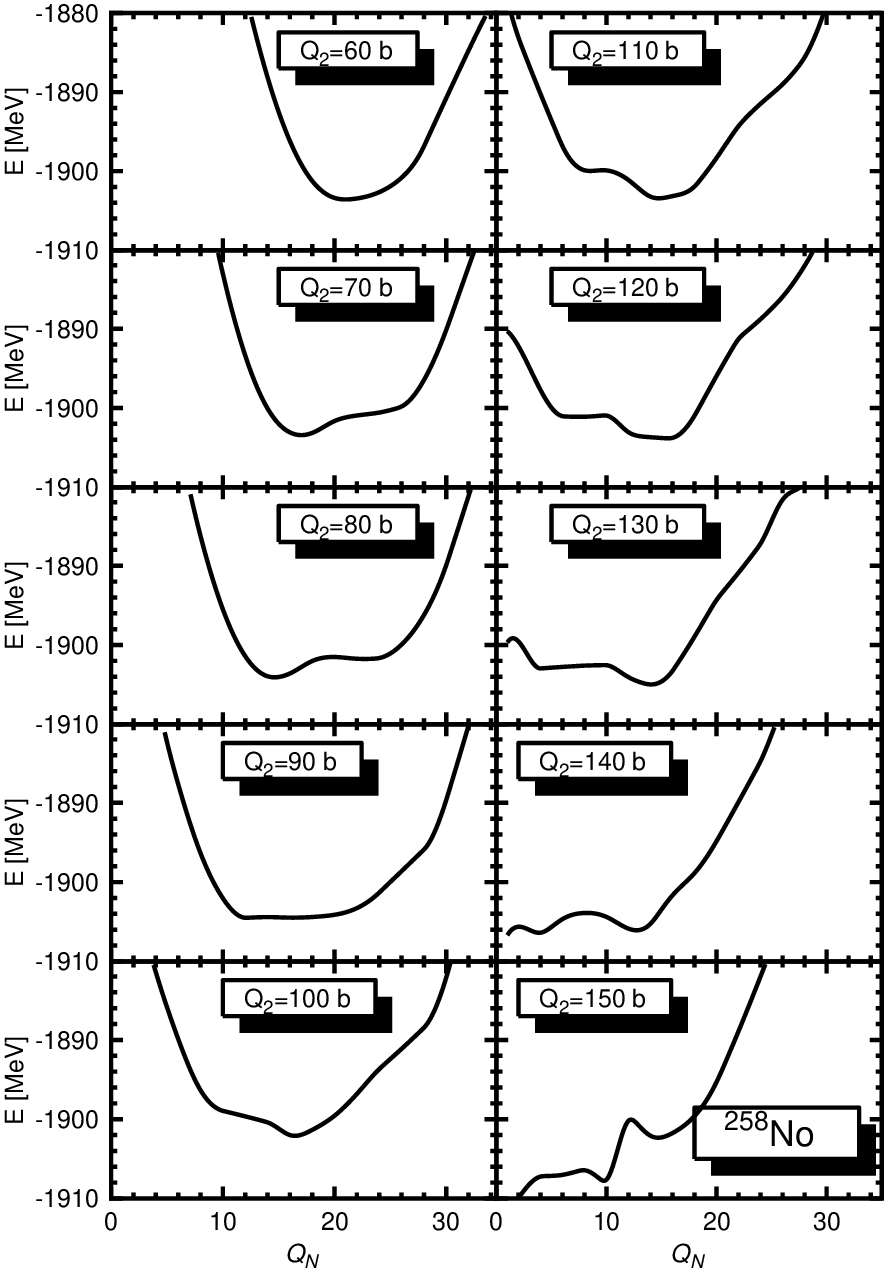, width=11cm, angle=00}
\end{center}
\caption{The same as in Fig.~5, but for the  $^{258}$No nucleus.}
\label{fig10}
\end{figure}

\pagebreak

\begin{figure}
\begin{center}
\leavevmode
\epsfig{file=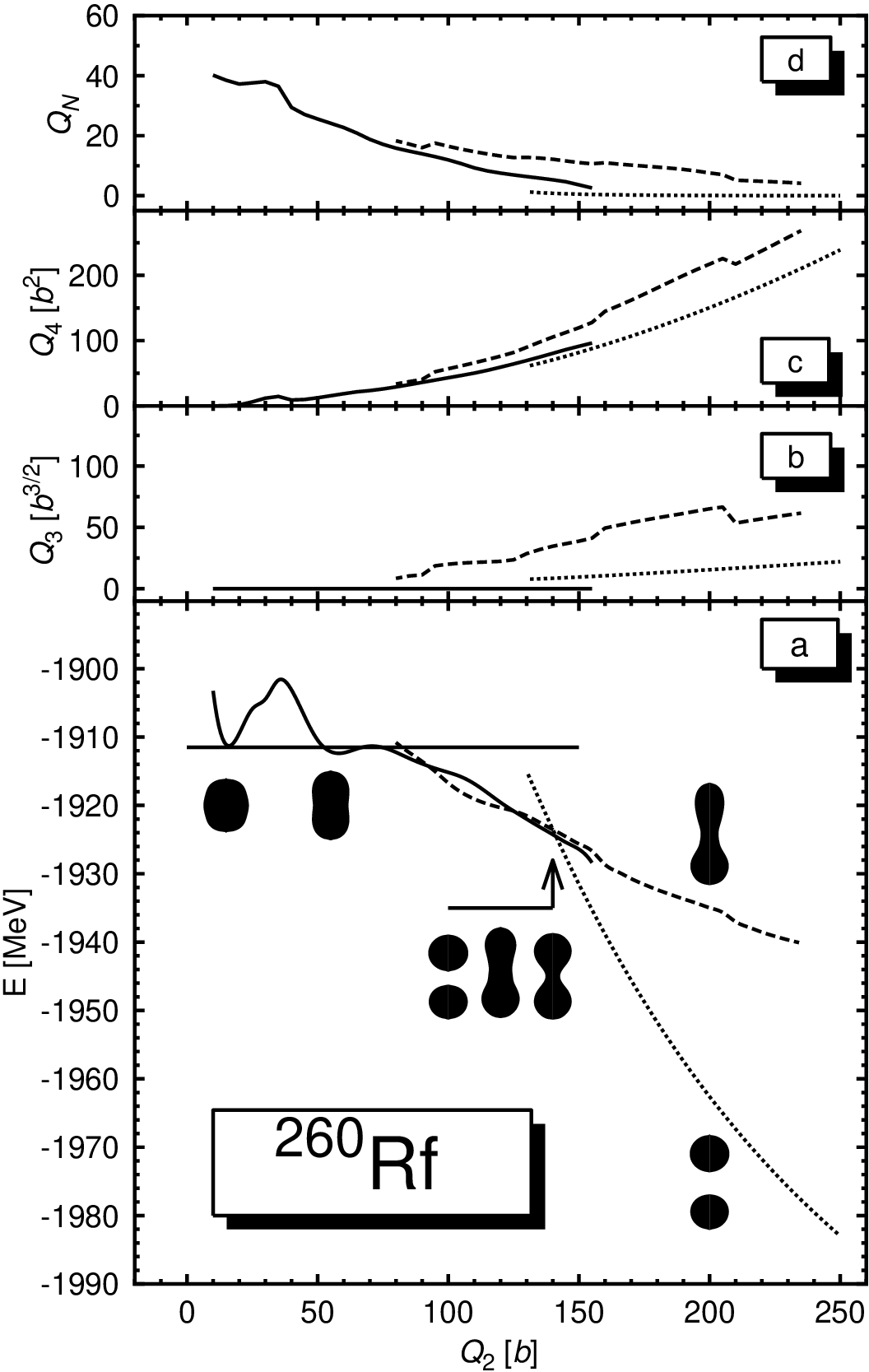, width=11cm, angle=00}
\end{center}
\caption{The same as in Fig.~9, but for the  $^{260}$Rf nucleus.}
\label{fig11}
\end{figure}

\pagebreak

\begin{figure}
\begin{center}
\leavevmode
\epsfig{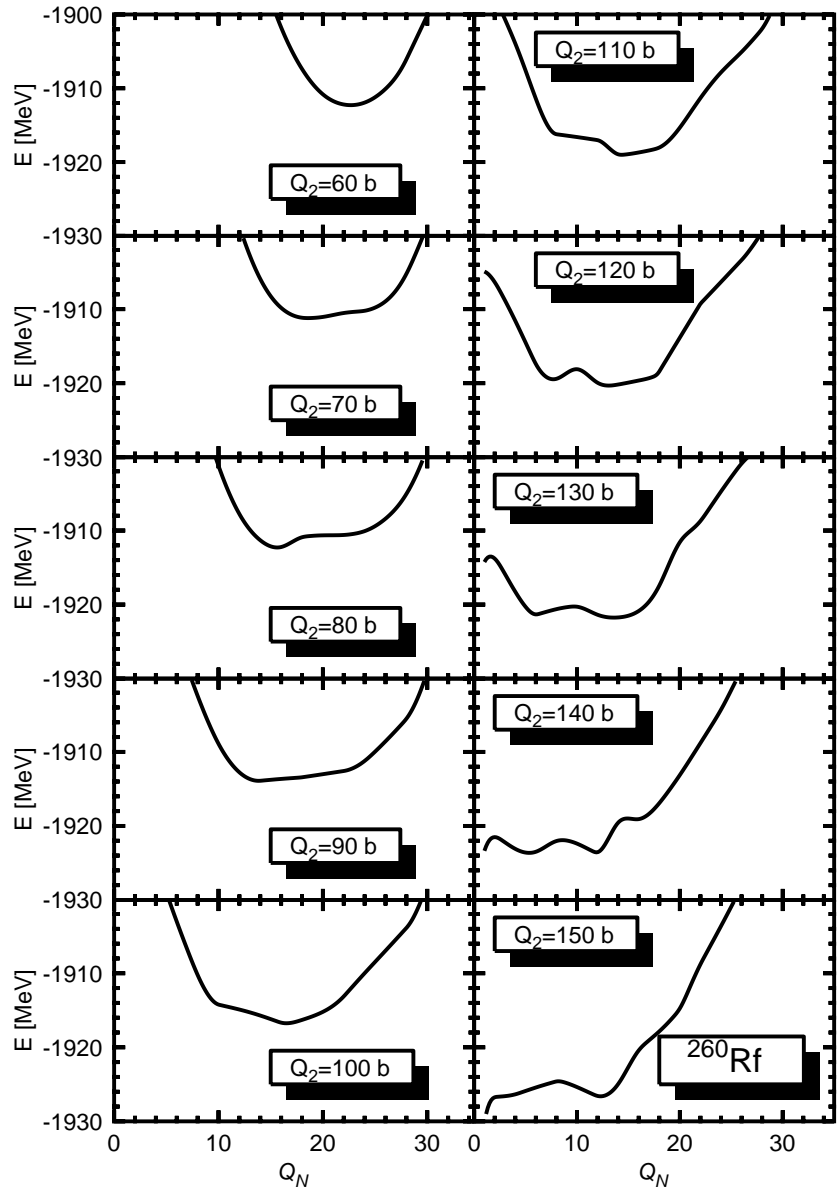}
\end{center}
\caption{The same as in Fig.~5, but for the  $^{260}$Rf nucleus.}
\label{fig12}
\end{figure}

\pagebreak

\begin{figure}
\begin{center}
\leavevmode
\epsfig{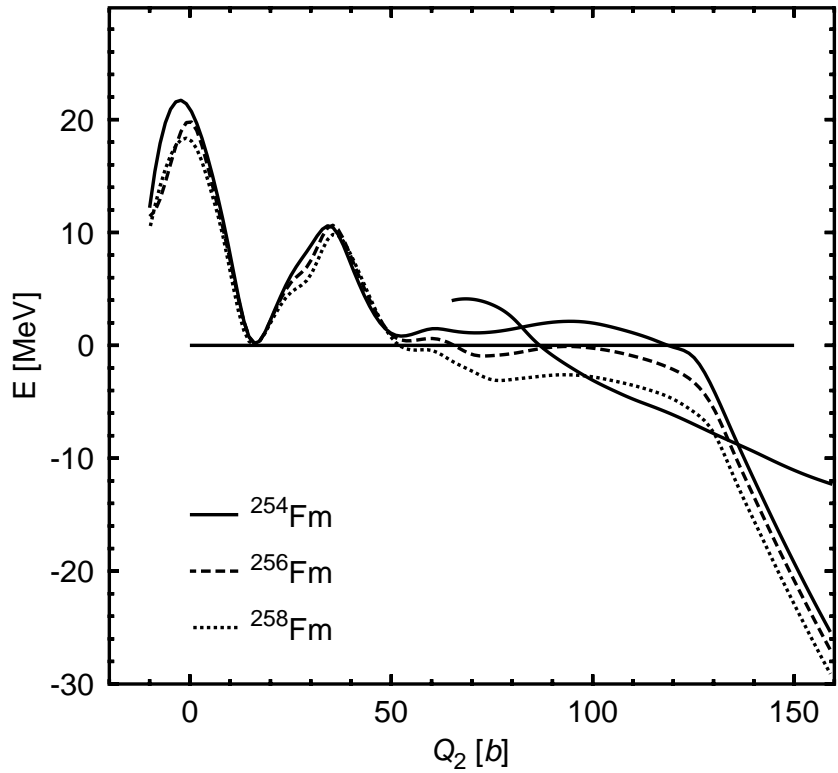}
\end{center}
\caption{The fission barriers for $^{254}$Fm, $^{256}$Fm and $^{258}$Fm 
         evaluated along the CF path. The ground state is set to zero 
         in the three cases. The second full line starting at $Q_2=60$ b 
	 correspond to the EF path of $^{254}$Fm.} 
\label{fig13}
\end{figure}

\pagebreak

\begin{figure}
\begin{center}
\leavevmode
\epsfig{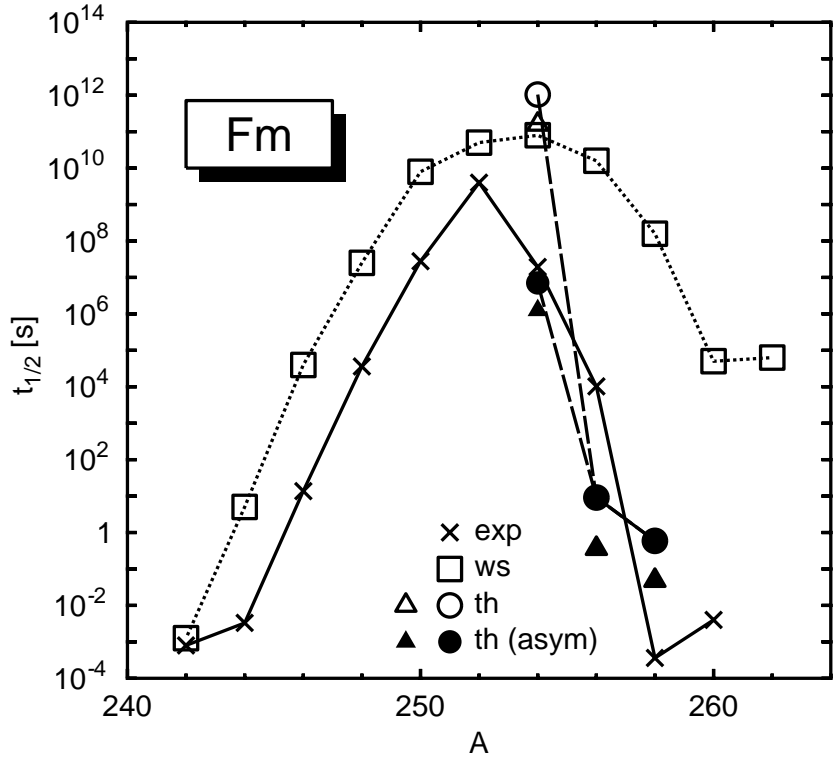}
\end{center}
\caption{The spontaneous fission half-life times of Fm isotopes as a function  
         of the mass number. The experimental data (exp.) are taken from 
         the NuDat data base while the theoretical estimates computed with a 
         model based on the Woods-Saxon (WS) potential  are taken from 
         Ref.\protect\cite{Sta99}. 
         The present estimates are represented by full circles for fission
	 along the CF path whereas the results for fission along the 
	 EF path are represented by open circles (note that both estimates
	 coincide in $^ {256-258}$Fm, these results were obtained assuming
         the axial symmetry of fissioning nucleus. Similar estimates done with 
         inclusion of the nonaxial degrees of freedom are marked by 
         triangles.}
\label{fig14}
\end{figure}

\end{document}